\documentclass[letterhead]{article} % DO NOT CHANGE THIS
\usepackage{aaai25}  % DO NOT CHANGE THIS
\usepackage{times}  % DO NOT CHANGE THIS
\usepackage{helvet}  % DO NOT CHANGE THIS
\usepackage{courier}  % DO NOT CHANGE THIS
\usepackage[hyphens]{url}  % DO NOT CHANGE THIS
\usepackage{graphicx} % DO NOT CHANGE THIS
\usepackage{natbib}  % DO NOT CHANGE THIS AND DO NOT ADD ANY OPTIONS TO IT
\usepackage{caption} % DO NOT CHANGE THIS AND DO NOT ADD ANY OPTIONS TO IT
\usepackage{algorithm}
\usepackage{algorithmic}
\usepackage{tcolorbox}
\usepackage{amsmath} 
\usepackage{fancyhdr}
\fancypagestyle{firstpagehf}
{
   \fancyhf{}
   \fancyhead[HC]{The AAAI 2025 Workshop on Economics of Modern ML:
Markets, Incentives, and Generative AI (MarketsGenAI)}
   \fancyfoot[C]{\thepage}
}
\usepackage{newfloat}
\usepackage{listings}
\DeclareCaptionStyle{ruled}{labelfont=normalfont,labelsep=colon,strut=off} % DO NOT CHANGE THIS
\floatstyle{ruled}
\newfloat{listing}{tb}{lst}{}
\floatname{listing}{Listing}
\newtcolorbox{bidderbox}[1]{colback=blue!5!white,colframe=blue!75!black, title=#1}
\title{InfoBid: A Simulation Framework for Studying Information Disclosure in Auctions with Large Language Model-based Agents}
\author{
    Yue Yin
}
\affiliations{
    Independent Researcher, San Francisco \\ 
    yueyin.io@gmail.com \\

}

\usepackage{bibentry}
\begin{document}

\thispagestyle{firstpagehf}
\maketitle

\begin{abstract}
In online advertising systems, publishers often face a trade-off in information disclosure strategies: while disclosing more information can enhance efficiency by enabling optimal allocation of ad impressions, it may lose revenue potential by decreasing uncertainty among competing advertisers. Similar to other challenges in market design, understanding this trade-off is constrained by limited access to real-world data, leading researchers and practitioners to turn to simulation frameworks. The recent emergence of large language models (LLMs) offers a novel approach to simulations, providing human-like reasoning and adaptability without necessarily relying on explicit assumptions about agent behavior modeling. Despite their potential, existing frameworks have yet to integrate LLM-based agents for studying information asymmetry and signaling strategies, particularly in the context of auctions.
To address this gap, we introduce InfoBid, a flexible simulation framework that leverages LLM agents to examine the effects of information disclosure strategies in multi-agent auction settings. Using GPT-4o, we implemented simulations of second-price auctions with diverse information schemas. The results reveal key insights into how signaling influences strategic behavior and auction outcomes, which align with both economic and social learning theories. Through InfoBid, we hope to foster the use of LLMs as proxies for human economic and social agents in empirical studies, enhancing our understanding of their capabilities and limitations. This work bridges the gap between theoretical market designs and practical applications, advancing research in market simulations, information design, and agent-based reasoning while offering a valuable tool for exploring the dynamics of digital economies.
\end{abstract}


\section{Introduction}
Today, display advertising drives a multi-billion-dollar market where publishers like Google and Meta sell user impressions to advertisers such as Coca-Cola, Amazon, and Nike. These impressions are sold via real-time auctions, where advertisers (bidders) submit bids, and the publisher (auctioneer) allocates the impressions and collects payments based on the auction's outcome. A defining feature of such markets is information asymmetry: publishers possess rich information about users—such as browsing history, interests, and demographic information—and also rich information about advertisers, potentially enabling the publishers to infer advertisers’ specific targeting goals, budget and requirement on return. Advertisers, on the other hand, have limited visibility into the auction environment and must rely on signals provided by the publisher to make their bidding decisions. With such dynamics, the publisher may reveal partial details about a user’s characteristics or the level of competition in the auction. This deliberate disclosure of information, also known as \textit{signaling}, shapes advertisers’ expectations and bidding strategies. For example, advertisers selling luxury watches may place higher bids on keywords (i.e., signals) associated with affluent consumers, as these users align more closely with their target audience, representing higher potential value. By selectively revealing information, auctioneers can influence bidders’ perceptions and bidding strategies, which in turn impacts auction outcomes. This raises a central question: \textbf{What information should be disclosed to optimize auction outcomes, whether for the auctioneer revenue, bidder surplus, or overall social welfare?}

Theoretical research has provided foundational insights into market design and strategic behavior in advertising auctions, with contributions from both academic researchers, including Nobel laureates \cite{vickrey1961counterspeculation, myerson1981optimal}, and leading industry practitioners \cite{liu2021neural, ostrovsky2011reserve, he2021unified}. There is also a significant area of study focusing on information asymmetry and signaling strategy, exploring topics such as public vs. private information, equilibrium behavior under various information structures, the design of optimal signaling mechanisms, and algorithmic methods to derive these signals \cite{emek2014signaling, bergemann2022optimal, cheng2015mixtureselectionmechanismdesign, badanidiyuru2018targeting, milgrom1982theory, milgrom2010simplified, levin2010online, feinberg2005anonymous} . 

Despite these advances, a significant empirical gap remains, particularly in understanding how signaling and information asymmetry influence bidding behavior in practical settings. This challenge is not unique to information disclosure but extends to broader aspects of market design in online environments, characterized by sequential, interdependent interactions among multiple strategic agents, as seen in systems like advertising platforms and recommendation engines. Real-world validation is constrained by limited access to proprietary data and the high cost of online experiments, both in terms of operations and user experience impact. As a result, researchers have turned to simulation environments to bridge this gap, with notable frameworks developed by Amazon \cite{jeunen2023off}, eBay \cite{nguyen2023practical}, Uber \cite{ubersim} and Google \cite{ie2019recsim}, etc.

However, existing simulation frameworks face notable limitations. Many of these frameworks focus on bidding strategies, reserve price optimization, or auction formats, while lack the exploration of the effect of information disclosure strategies. Second, the simulated agents (bidders) are typically guided by pre-determined assumptions, i.e., their decision-making processes are parametrized with a limited set of predefined rules or structures. These assumptions may be encoded through heuristic models, structured stochastic processes, or parametrized frameworks such as neural networks. While these approaches provide tractable models of agent behavior, they often lack the richness and adaptability of human decision-making, as well as the interpretability of reasoning that characterizes rational human agents. 

The emergence of large language models (LLMs), such as GPT-4, offers a promising solution to the limitations of traditional agent-based models, particularly in addressing the challenges of adaptability and human-like reasoning \cite{openai2024gpt4technicalreport, feizi2023online, touvron2023llama, zeng2023glm130bopenbilingualpretrained}. Unlike conventional approaches that rely on parametrized assumptions, LLMs enable the creation of adaptive agents capable of reasoning with limited input assumptions, making nuanced decisions, and providing interpretable explanations for their actions. Early applications of LLM-based agents in simulation frameworks have demonstrated such frameworks' potential to enhance the realism and interpretability of competitive environments, including auctions \cite{chen2023put, zhaocompeteai}. However, existing frameworks have not yet incorporated information disclosure strategies into LLM-driven simulations, particularly in the context of online advertising auctions. This represents a significant gap, as understanding the interplay between signaling strategies and agent behavior is crucial for advancing both theoretical insights and practical applications in auction design.

In this paper, we address the critical gap in understanding the role of information disclosure in auctions by introducing \textbf{InfoBid}, a simulation framework that leverages LLM-based agents (using GPT-4o) to study strategic behavior under information asymmetry in multi-agent auctions. A critical focus of this study is to see whether these LLM-based agents exhibit rational behavior, relying on their inherent reasoning capabilities rather than pre-parameterized assumptions or explicitly encoded decision-making processes. Therefore, our experiments are designed to provide minimal public information and avoid imposing explicit assumptions about agents' internal models. Instead, agents are tasked with reasoning based on disclosed information and their own knowledge. Within this setup, we systematically explore diverse private signaling strategies, ranging from full disclosure, which reveals precise valuations of an advertisement, to selective pooling, which aggregates and shares approximate signals, to randomized pooling, which probabilistically discloses information. These strategies allow us to analyze how information impacts key auction outcomes—such as revenue, bidder surplus, and social welfare—and to observe the reasoning and adaptability of LLM-based agents under varying scenarios.

\subsection{Findings} 
Our key findings after running a few simulations under this framework are the following:
\begin{itemize}
    \item \textbf{Rationality Analysis on Agent's Decision-making}: We find that bidder behavior reflects both alignment with and deviations from truthful bidding under various signaling strategies. Specifically, revealing relative positions disrupts truthful bidding, as bidders adjust their bids based on perceived advantage in the competition. However, our results suggest that adding richer information, such as the average true valuation within tiers, mitigates these deviations and stabilizes bidding. These findings echo social learning theories, such as social comparison and anchoring effects.

\item \textbf{Agents' Reasoning}: Our analysis of bidders’ explanations reveals that LLM-based agents do not explicitly consider competitors’ values or strategies in their decision-making. While this behavior aligns with the theoretical expectation in second-price auctions, where truthful bidding is the dominant strategy, we don't know whether these outcomes stem from the agents' trained rationality or limitations in perceiving competitive dynamics. Exploring these aspects in alternative auction formats, such as first-price auctions, could provide valuable insights into their reasoning capabilities and adaptability.

\item \textbf{Revenue and Social Welfare Analysis}: Our results indicate that Pool-High strategies targeting high percentiles and incorporating average true values in pooled information yield the highest auctioneer revenue. This aligns with theoretical insights from economic literature, as the strategy intensifies competition among top-tier bidders and extract high information rents, driving revenue gains. Additionally, we find that social welfare improves with increased information disclosure to high-value bidders, underscoring the importance of transparency in achieving efficient resource allocation. Together, these findings suggest that carefully designed signaling strategies can simultaneously enhance auctioneer revenue and social welfare outcomes.

\end{itemize}

\subsection{Contributions} The contributions of this paper are three-fold:
\begin{enumerate}
    \item Framework for LLM-Based Information Strategies: We pioneer a simulation framework that integrates information disclosure strategies into LLM-based agents, offering a novel testbed for studying auction mechanism in online advertising markets.

\item Development of a Simulated Environment: The framework provides a flexible and scalable environment to explore various auction setups, information structures, and agent behaviors, bridging the gap between theoretical insights and empirical validation.

\item Novel Insights into Agent Behavior and Economic Theory: Our study reveals how LLM agents reason and adapt to information asymmetry, producing behaviors that align with auction theory. These findings not only validate the utility of LLM-based simulations but also guide future research on agent behavior, mechanism design, and market efficiency.
\end{enumerate}

\section{Methodology}
The \textbf{InfoBid} framework is an extensible simulation environment designed to study the impact of information disclosure strategies on agents' strategic behavior and auction outcomes. The framework comprises several key components: \textbf{Agents}, which include the auctioneer and multiple bidders, each with distinct roles and decision-making capabilities; \textbf{Information Structure}, which specifies the information to be disclosed, pooled, or hidden, shaping the strategic dynamics of the auction; and the \textbf{Auction and Signaling Process}, which manages the flow of information and bids between agents, simulating real-world auction interactions. These components, together with \textbf{configurable simulation parameters}, make InfoBid a versatile testbed for exploring the relationships between auction design, information asymmetry, and strategic decision-making. Additionally, InfoBid supports the creation and evaluation of diverse information disclosure policies, providing a scalable and flexible platform for advancing research in market simulations and auction outcome.

\begin{figure}[t]
\centering
\includegraphics[width=0.9\columnwidth]{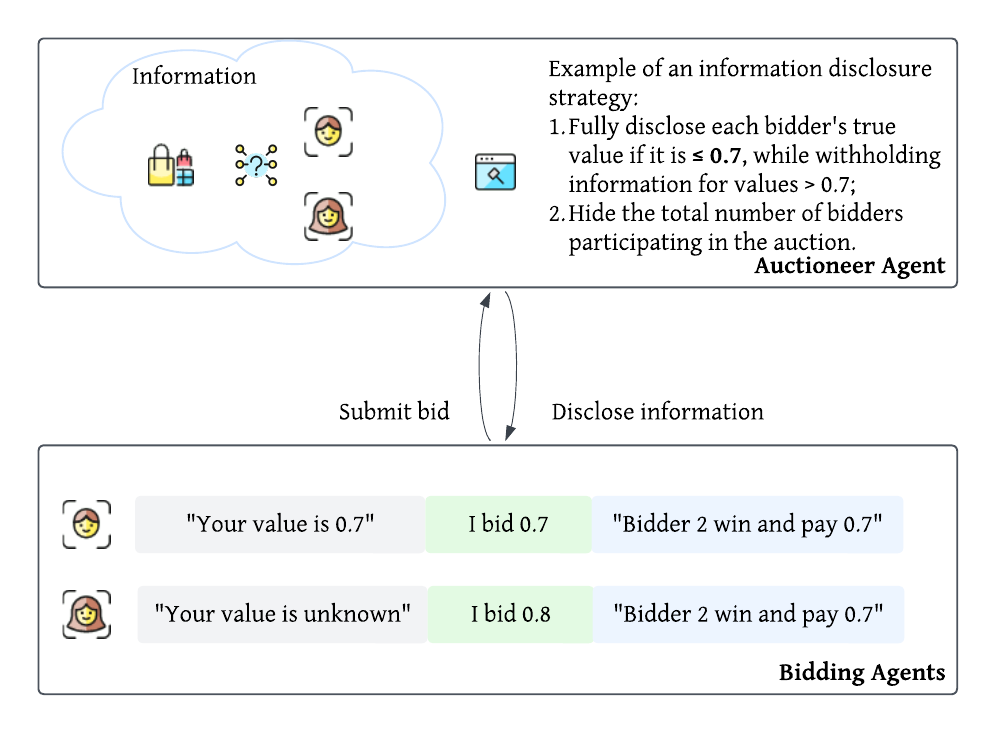} % Reduce the figure size so that it is slightly narrower than the column. Don't use precise values for figure width.This setup will avoid overfull boxes.
\caption{\textbf{Information flow between auctioneer and bidders in an example auction.} Gray-colored messages represent private information, such as \textit{Your value is 0.7} provided individually to bidders by the auctioneer. Green-colored text represents sealed bids submitted privately by bidders to the auctioneer, such as \textit{I bid 0.7} or \textit{I bid 0.8}. Blue-colored messages signify public information, such as the auctioneer announcing, \textit{Bidder 2 wins and pays 0.7}, visible to all participants. This example showcases a specific disclosure strategy, including conditions for revealing true values and withholding total bidder count.}
\label{fig-flow}
\end{figure}

\subsection{Agent Design}

\subsubsection{Auctioneer}

The auctioneer serves as the central entity responsible for managing the auction process and typically holds more information than the bidders. Two settings are widely studied in the literature \cite{emek2014signaling}:
\begin{itemize}
    \item \textbf{Known-Valuations}: In this setting, the auctioneer possesses complete knowledge of all bidders’ true valuations for the auctioned item. These valuations are often derived from detailed item characteristics or bidder profiles that the auctioneer has exclusive access to. For example, in online advertising, the auctioneer (e.g., the publisher) might have accurate predictions of how much each advertiser values an impression based on user demographics, historical click-through rates, or conversion likelihoods. With full knowledge of bidder valuations, the auctioneer can precisely tailor disclosure strategies—such as revealing exact values, pooling certain groups, or selectively sharing information—to maximize specific objectives, such as auctioneer revenue or social welfare. 
    \item \textbf{Bayesian Setting}: The auctioneer does not know the exact valuations of the bidders but operates with a prior belief about these valuations. For instance, the auctioneer might know general characteristics of the item being auctioned but be uncertain about how different types of bidders—such as cost-sensitive or quality-focused participants—value the item. These bidder types are not directly observable to the auctioneer, introducing an additional layer of complexity. 
\end{itemize}

\subsubsection{Bidder}

Bidder agents represent the participants in the auction and are tasked with formulating strategic bids based on the signals received from the auctioneer and their own knowledge. Unlike the auctioneer, bidders lack information about the auctioned item and are unaware of their true valuations. In this study, we employ GPT-4o as the reasoning engine for bidders, with the temperature set to 0 to ensure reproducibility. Bidders generate their bids in two steps: (1) estimate their true valuations based on the provided signals and their own knowledge. For instance, a signal such as \textit{"You are among the high value bidders"} may lead a bidder to infer a relatively high valuation while remaining uncertain about its precise value; (2) formulate their bids based on these estimated values. For example, a bidder estimating their valuation at 0.75 might bid slightly below this value to account for uncertainty while remaining competitive. Additionally, bidders are required to explain their bidding decisions.

\subsection{Information Structure}
The information structure in our framework is designed to capture the strategic flow of data between the auctioneer and bidders, highlighting the interplay between shared and individualized information. It is categorized as follows:
\begin{itemize}
    \item \textbf{Public Information}: Public information is shared among all agents (including the auctioneer and all bidders) and serves as a common reference point for their decision-making. In this study, public information may include:
    \begin{itemize}
        \item Auction Rules: The mechanism governing the auction, such as whether it is a first-price or second-price auction, the number of bidders, and whether ties are broken randomly or systematically; 
        \item Common Prior on Bidders' Distribution: For example, bidders and the auctioneer may agree that valuations are uniformly distributed over $[0, 1]$. Also, if the environment assumes bidders have different types (e.g., "high-budget" vs. "low-budget" bidders), the prior distribution of these types may also be common knowledge; 
        \item Public Knowledge of the Signaling Map: The signaling map describes how realized random states (e.g., bidders' valuations) are transformed into signals disclosed to bidders. For example, the signaling map might specify that high-value bidders (e.g., those above the 80th percentile) will have their information pooled, while low-value bidders (e.g., those below the 80th percentile) will have their valuations fully disclosed. This map may be publicly known and shapes the bidders' expectations of the auction's informational context.
        \item Auction outcome: After each round, the auctioneer reveals the auction results, which are accessible to all bidders. This includes the name or identifier of the bidder who won the item and The final price paid for the item, typically determined by the second-highest bid in a second-price auction.
    \end{itemize}
    \item \textbf{Private Information}: Private information is exclusive to each individual bidder and remains hidden from others, allowing the auctioneer to strategically influence personalized bidding behavior through tailored signals. In our framework, private information pertains primarily to a bidder’s valuation of the auctioned item. The auctioneer may disclose this information using various signaling strategies, including:
    \begin{itemize}
            \item Full Disclosure: The auctioneer reveals the exact value of a bidder’s true valuation. For example, a bidder might be informed, \textit{"Your valuation for this item is 0.7."} This approach eliminates uncertainty for the bidder, encouraging truthful bidding behavior.
            \item Selective pooling involves aggregating information and sharing approximate signals with specific groups of bidders, allowing the auctioneer to influence their bidding strategies indirectly. For instance, tiered pooling might provide relative placement information, such as stating, “You are among the top value bidders” or “Your value falls into the low tier,” rather than revealing precise valuations. Alternatively, randomized pooling introduces an additional layer of uncertainty by selectively disclosing pooled information to certain bidders without a deterministic pattern. 
    \end{itemize}
\end{itemize}

\subsection{Auction and Signaling Process}
\textbf{InfoBid} employs a structured six-phase pipeline, from initializing the auction context to announce the auction outcome, to simulate the auction and signaling process, providing a comprehensive approach to analyzing the effects of information disclosure strategies on auction outcomes and bidder behavior. Detailed introduction of the phases is in Appendix. 

\section{Experiments}
\subsection{Simulation Design}

In this study, we focus on the \textbf{Known-Valuation} setting, where the auctioneer has complete knowledge of each bidder's true valuation for the auctioned item. For simplicity, the experiments include 10 bidders, with each bidder’s valuation for a given item independently drawn from a uniform distribution over $[0, 1]$.

We adopt a sealed-bid second-price auction without a reserve price—a widely recognized auction format used in both theoretical and applied contexts \cite{vickrey1961counterspeculation}. In this mechanism, bidders submit private bids without knowledge of others’ bids. The bidder with the highest bid wins the item but pays the second-highest bid. This format incentivizes bidders to bid \textit{truthfully} according to their perceived valuations, providing a robust framework to analyze how various information disclosure strategies influence strategic decision-making.

By default, the auctioneer publicly shares the following information to establish a common context for all participants:
\begin{enumerate}
    \item The auction type (sealed-bid second-price) and the number of bidders.
    \item The common prior that all bidders’ valuations for each item are independently drawn from a uniform distribution over $[0, 1]$.
\end{enumerate}
The experiments systematically vary the \textit{private signaling strategies} used to communicate information about bidders’ valuations. Importantly, these signaling strategies are not shared as public knowledge, meaning bidders must rely solely on the default public information, the private signals they receive, and any implicit reasoning capabilities derived from their pretrained knowledge. They do not have explicit knowledge of the information other bidders receive. 
The experimented private signaling strategies include the following (For further details on the prompts used in each scenario, refer to the Appendix)
\begin{itemize}
    \item Full disclosure: bidders receive their exact valuation;
    \item Tiered pooling ("Pool-High" and "Pool-Low"): bidders are grouped into quantile-based tiers. For instance, the "Pool-High" strategy provides pooled information to bidders whose true valuations fall within the higher percentiles while fully disclosing valuations to the remaining bidders. The mathematical formulation of "Pool-High" is shown in Eq. \ref{eq1}, where $F(v_i)$ is the cumulative distribution function (CDF) of $v_i$, $q$ is the quantile threshold, and $p(v_i)$ represents the pooled information. Similarly, the Pool-Low strategy applies pooling to lower-valued bidders, selectively revealing their positional context. Tiered pooling includes two variants of pooled information: 1) Tier-only information, where bidders are informed solely of their relative position within a tier, e.g., “You are among the high tier of bidders."; 2) Tier and average value information, where bidders receive their relative position along with the average true valuation within their tier, e.g., “You are among the high tier of bidders, and the average valuation in your tier is 0.8.”
    \begin{align}
s(i)\begin{cases}
    v_i & F(v_i) < q, \\ 
p(v_i) & F(v_i) \geq q
\end{cases}
\label{eq1}
\end{align}

    \item Randomized pooling: randomly pooling bidder information and selectively disclosing it based on predefined probabilities (Eq. \ref{eq2})
\begin{align}
s(i) &= 
\begin{cases} 
v_i & \text{with probability } q, \\
p(v_i) & \text{with probability } 1 - q,
\end{cases}
\label{eq2}
\end{align}

\end{itemize}
 
Both tiered pooling and randomized pooling are parameterized using a set of probalistic threshold ($q$ in the equation). These thresholds control either the quantile boundaries for tiered pooling or the probability of information disclosure in randomized pooling, enabling direct comparisons between the two approaches.

To ensure robust findings, each simulation configuration is executed over 100 independent rounds. In each round, a new item is introduced, and bidders are assigned fresh valuations drawn from the uniform distribution, ensuring independent and identically distributed (i.i.d.) conditions across rounds. For simplicity, the framework does not incorporate iterative learning across rounds, treating each round as an isolated snapshot of decision-making process. See Appendix for a list of simulation variants configurations in our study.

\subsection{Evaluation Metrics}
To analyze the effectiveness of information disclosure strategies and bidding behaviors, we evaluate the following metrics for each configuration:

\begin{itemize}
   
     \item \textbf{Bid Deviation from Rational Behavior:} In a second-price sealed-bid auction, rational behavior is defined as bidding truthfully, i.e., submitting a bid equal to the bidder’s estimated true value. This metric evaluates the extent to which submitted bids align with the bidder’s estimated valuation. Bid deviation is quantified as the percentage of cases where a bidder’s submitted bid is greater than, equal to, or less than their estimated value, providing insights into deviations from theoretically optimal bidding behavior.

    \item \textbf{Revenue:} Revenue is calculated as the price paid by the winning bidder in each round, summed over all rounds in the simulation. This metric provides insight into the auctioneer's performance under different disclosure strategies.

    \item \textbf{Social Welfare:} Social welfare is defined as the sum of the true valuations of the winning bidders across all rounds. Mathematically, it can be expressed as:
    \[{Social Welfare} = \sum_{r=1}^{R} v_r\]
    where \( R \) is the total number of auction rounds, and \( v_r \) represents the true valuation of the winning bidder in round \( r \). This metric evaluates the overall efficiency of the auction in allocating items to bidders with the highest valuations.
    
\end{itemize}

\subsection{Results}
\subsubsection{Bidder's Rationality Analysis}
\begin{figure}[t]
\centering
\includegraphics[width=0.9\columnwidth]{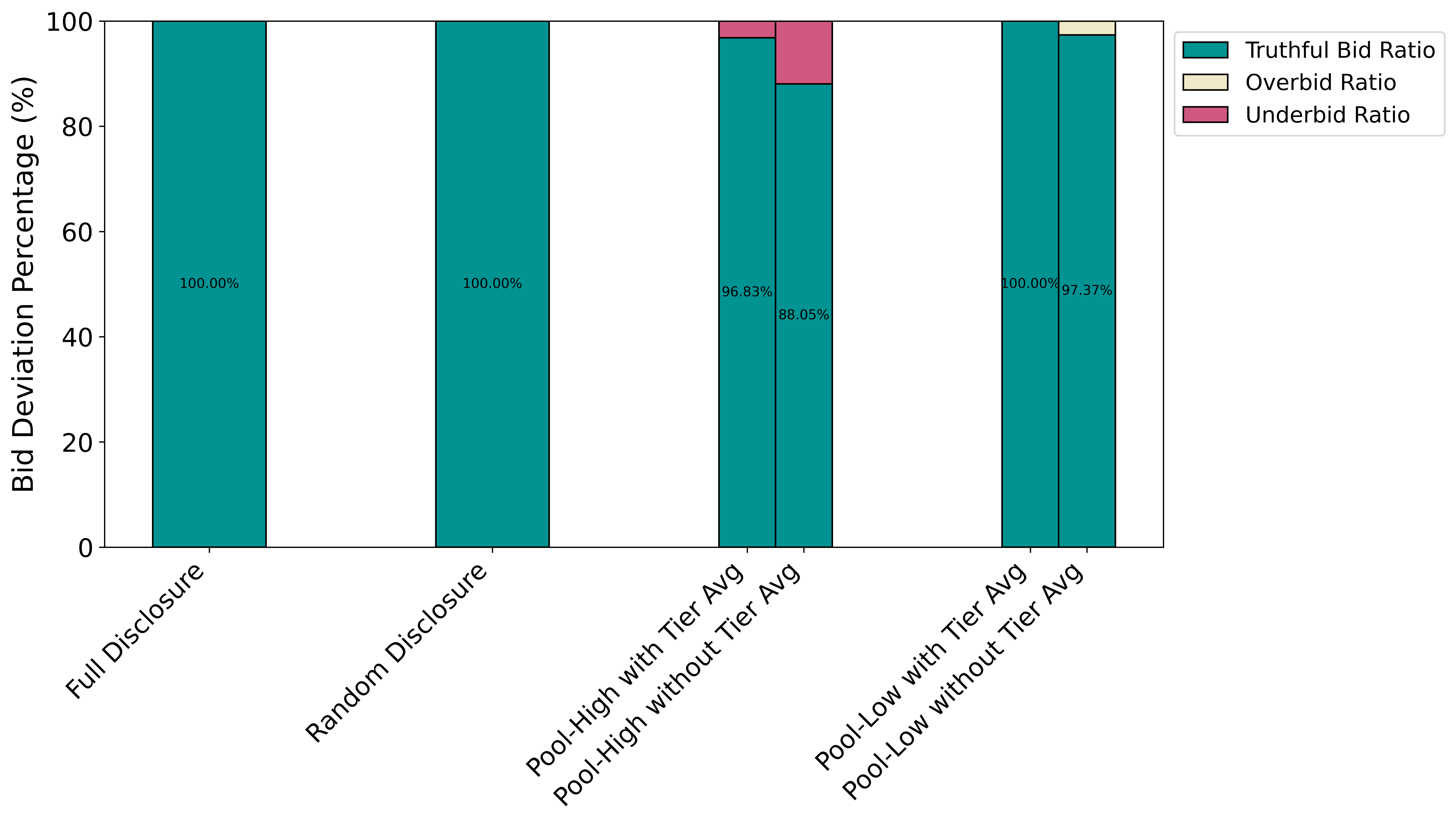} % Reduce the figure size so that it is slightly narrower than the column. Don't use precise values for figure width.This setup will avoid overfull boxes.
\caption{\textbf{Impact of Signaling Strategies on Bid Deviation Behavior} The chart illustrates the percentage of deviated bidding behavior across various signaling strategies, as represented on the x-axis. The y-axis denotes the proportion of bids categorized into three groups: truthful bidding (where the bid value equals the estimated value), overbidding (where the bid value exceeds the estimated value), and underbidding (where the bid value is lower than the estimated value). Each bar is derived from 4000 bid records. }
\label{fig-bid}
\end{figure}
\begin{figure*}[t]
\centering
\includegraphics[width=0.9\textwidth]{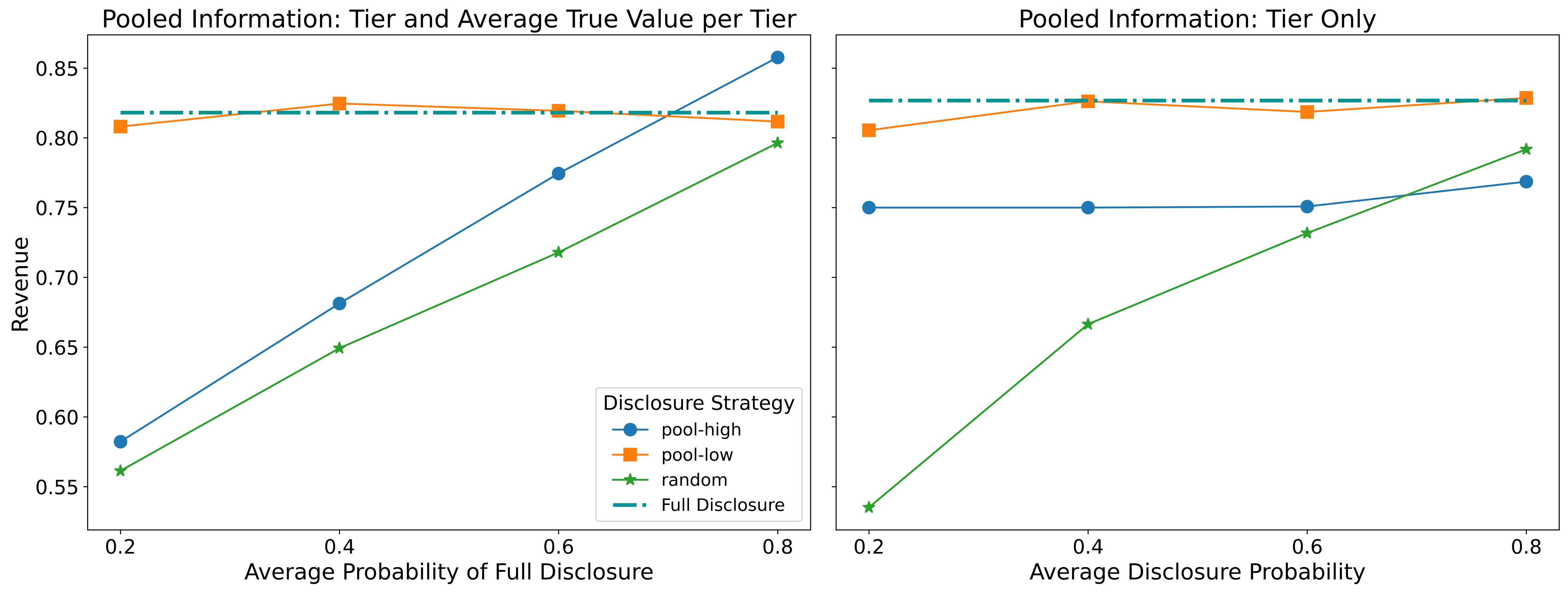} % Reduce the figure size so that it is slightly narrower than the column. Don't use precise values for figure width.This setup will avoid overfull boxes.
\caption{\textbf{Revenue Comparison Across Disclosure Strategies and Scenarios.} The x-axis represents the average probability of full disclosure under various scenarios. For instance, pooling bidders with values above the 20th percentile corresponds to a full disclosure probability of 0.2. The y-axis shows the average revenue generated per round for each configuration. The chart compares four disclosure strategies: \textbf{Full Disclosure} (depicted as a horizontal line); \textbf{Pool-High} and \textbf{Pool-Low} with each strategy evaluated across different cut-off values; \textbf{Random}, where bidders are randomly selected for full disclosure without providing additional information to the remaining bidders.}
\label{fig-revenue}
\end{figure*}

The analysis of rationality in this study can be divided into two aspects: (1) the rational inference of valuations, where bidders estimate their value based on disclosed information, and (2) the rational bidding behavior, defined as submitting bids that align with the inferred valuations. While theoretical equilibrium prescribes truthful bidding—submitting bids equal to the estimated valuation—agents may deviate depending on the disclosed information and their interpretation of it.

First, we quantitatively analyze the extent to which bidders' submitted bids align with their inferred valuations as illustrated by figure \ref{fig-bid}. Under full information, bidders consistently engage in truthful bidding, submitting bids precisely equal to their estimated valuations. This outcome validates the rational behavior of LLM agents' ability to reason logically and bid optimally when provided with unambiguous and complete information. Similarly, in randomized pooling—where signals lack relative standing or positional details but indicate whether a value is pooled or disclosed—truthful bidding also achieves 100\% consistency. The absence of comparative information simplifies decision-making, as agents cannot infer additional competitive nuances, leading them to align their bids with their estimated values.

In contrast, tiered information disclosure disrupts this equilibrium. Under the "Pool-High" strategy—where high-value bidders are pooled and informed of their relative position—bidders tend to underbid. Conversely, the "Pool-Low" strategy, which pools low-value bidders and informs them of their tier, prompts overbidding. These behaviors align with social comparison theory, which suggests that individuals adjust their actions based on their perceived position relative to others\cite{tor2023neuroscience, garcia2013psychology}. High-value bidders in the "Pool-High" tier, aware of the intense competition within their group, may bid conservatively to avoid overpaying. Conversely, low-value bidders in the "Pool-Low" tier may perceive an opportunity to compete more aggressively, resulting in overbidding. 
\begin{figure*}[t]
\centering
\includegraphics[width=0.9\textwidth]{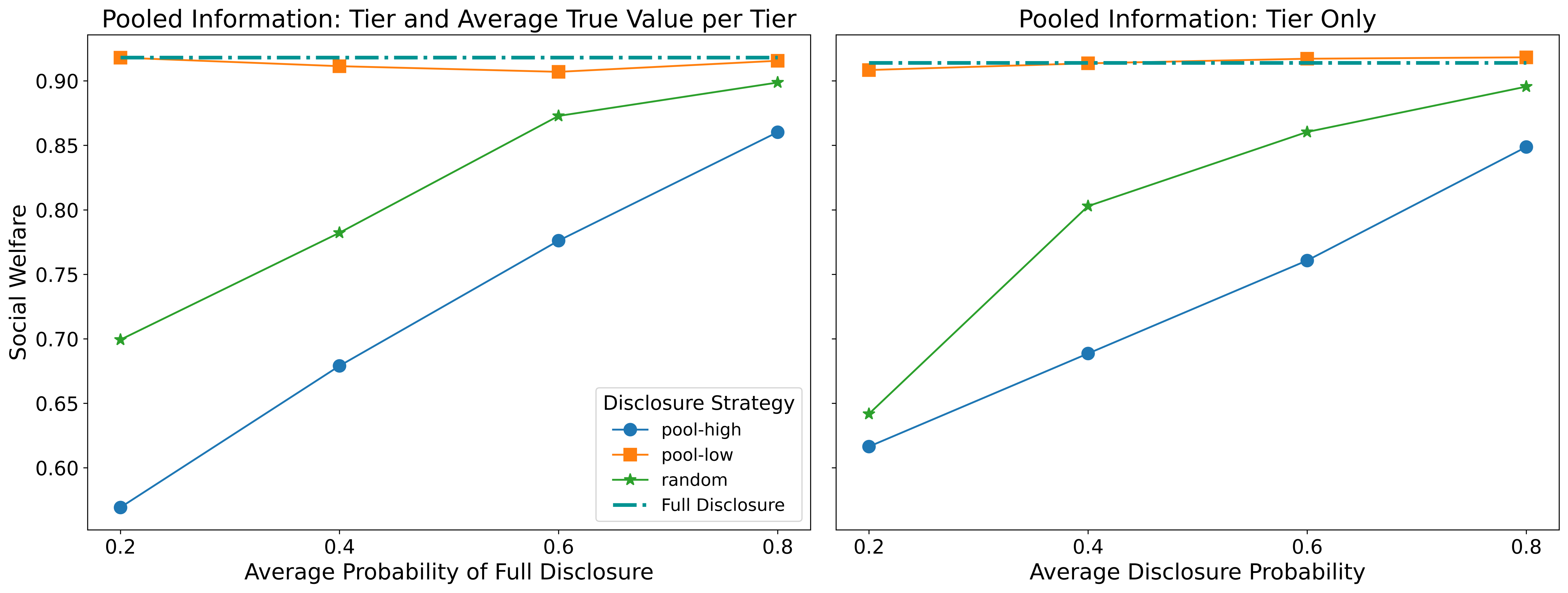} % Reduce the figure size so that it is slightly narrower than the column. Don't use precise values for figure width.This setup will avoid overfull boxes.
\caption{\textbf{Social welfare Comparison Across Disclosure Strategies and Scenarios.} The x-axis represents the average probability of full disclosure under various scenarios. For instance, pooling bidders with values above the 20th percentile corresponds to a full disclosure probability of 0.2. The y-axis shows the average revenue generated per round for each configuration. The chart compares four disclosure strategies: \textbf{Full Disclosure} (depicted as a horizontal line); \textbf{Pool-High} and \textbf{Pool-Low} with each strategy evaluated across different cut-off values; \textbf{Random}, where bidders are randomly selected for full disclosure without providing additional information to the remaining bidders.}
\label{fig-social}
\end{figure*}

The inclusion of additional value information, such as the average true valuation within the pooled tier, further reduces deviations from truthful bidding. This richer information instills greater confidence in bidders’ valuations, resulting in more stable and consistent bidding behavior. Analyzing bidding records reveals numerous instances where agents update their estimated valuations to align with the provided average value and subsequently bid accordingly (see the following example). This behavior reflects the anchoring effect \cite{tversky1982judgment}, a cognitive bias in which individuals rely heavily on reference points—such as disclosed average values—when making decisions. By anchoring their bids to these provided values, bidders demonstrate a clear reliance on structured information to guide their strategies.

Second, we delve into bidders' explanations (see Appendix for examples) to understand how their beliefs are updated based on accessible information. Interestingly, the records reveal no references to the values or potential behaviors of other bidders (as illustrated in the example below). This suggests that the LLM-based agents' decision-making processes do not explicitly incorporate strategic reasoning about competitors. Notably, in the context of second-price auctions, this behavior aligns with theoretical expectations, as bidding truthfully to one’s estimated value is the dominant strategy and should not be influenced by the potential actions or valuations of competitors.

\begin{tcolorbox} \textit{Given that my true value is in the high-value tier and the average value of bidders in this tier is 0.6657625810811798, I estimate my true value to be around this average. In a second-price auction, the optimal strategy is to bid your true value. Therefore, I bid 0.6657625810811798, aligning with my estimated true value.} \end{tcolorbox}

However, it remains unclear whether this observed behavior is a result of the trained rationality of LLM agents or due to a limitation in their reasoning capabilities that prevents them from factoring competition into their decision-making. We did not extensively analyze this in the present study, but it presents an interesting avenue for future research. For instance, shifting to a first-price auction, where strategic reasoning about competitors is critical, could provide further insights into the agents' capabilities and the impact of different auction formats on their decision-making processes.

%Table \ref{tab:llm_rationality_baseline} 
%\begin{table}[ht]
%\centering
%\begin{tabular}{l|l|l}
%\textbf{Number of Bidders} & \textbf{Disclose} & \textbf{Hide} \\ \hline
%5                         & 100\%                              & 100\%                            \\ 
%10                         & 100\%                              & 100\%                            \\ 

%\end{tabular}
%\caption{Percentage of Bids Matching True Valuations under Full Disclosure}
%\label{tab:llm_rationality_baseline}
%\end{table}

\subsubsection{Revenue analysis}
The revenue results, illustrated in Figure \ref{fig-social}, reveal distinct patterns across the different information disclosure strategies. The Pool-Low strategy and Full Disclosure consistently generate similar revenue outcomes. This similarity arises because both strategies effectively reveal the true valuations of the highest bidders. As long as these bidders bid truthfully, they secure the winning bid, resulting in comparable revenue levels across these strategies.

In contrast, the Pool-High strategy exhibits increasing revenue as the percentage of disclosed information rises. This effect stems from the likelihood that top bidders will more accurately estimate their true valuations with additional information. As we see in the rationality analysis, without full disclosure, high-value bidders often rely on pooled signals, leading to conservative value estimations. As more information is disclosed, their bid accuracy improves, driving up revenue.

When pooled information includes only the tier, bidding records suggest that high-value bidders typically estimate their bids a stable constant (our bidding records suggest 0.75, which happens to be the average value from a uniform distribution drawing $[0.5,1]$), resulting in stable revenue across quantiles. However, when pooled information includes both the tier and the average true value within the tier, the additional detail creates significant information rents for top-tier bidders. For instance, pooling the top two bidders yields a higher average true value than pooling the top five bidders because the smaller subset is concentrated at the upper end of the value distribution. As we see from the rationality analysis, the higher average shapes bidder's belief and incentivizes aggressive bidding among top-tier bidders, leading to revenue gains. These findings align with theoretical results from \citeauthor{bergemann2022optimal}, which demonstrate that strategically designed information disclosure—such as pooling and signaling—can generate higher revenue than full disclosure by effectively balancing uncertainty and competition.

\subsubsection{Social Welfare Analysis}
The social welfare results, also shown in Figure \ref{fig-social}, provide insights into how information disclosure strategies influence the overall efficiency of the auction system. Social welfare, defined as the total true value obtained by the winning bidders, serves as a critical metric for evaluating resource allocation efficiency—ensuring that items are awarded to those who value them most.

The Pool-High strategy consistently underperforms compared to other strategies at lower thresholds. This underperformance is due to the reduced transparency caused by pooling information for high-value bidders, which limits their ability to bid effectively, thereby reducing allocation efficiency. However, as the threshold increases (e.g., 0.8), the transparency improves, and social welfare outcomes under Pool-High approach those of Full Disclosure, reflecting more efficient allocation. Similarly, for the Random-pooling strategies, social welfare increases with the probability of full disclosure. 

Conversely, the Full Disclosure and Pool-Low strategies consistently yield higher social welfare by ensuring that high-value bidders—who contribute most significantly to social welfare—receive adequate information. The stability of social welfare under Pool-Low across thresholds indicates that pooling information for lower-value bidders has minimal adverse impact, provided high-value bidders are sufficiently informed.

\section{Conclusion}

In this paper, we propose InfoBid, a simulation framework leveraging large language models (LLMs) to explore the effects of information disclosure strategies in auction environments. Through simulations with diverse signaling strategies, we demonstrate the critical role of targeted information disclosure to selective bidders in shaping auction outcomes. Additionally, our analysis highlights the importance of examining agents' reasoning processes to uncover their capabilities and limitations. By aligning with both economic and social learning theories, our findings offer valuable insights into optimizing auction outcomes through strategic information disclosure.  This study advocates for future research to further understand and enhance LLM agents’ strategic reasoning, advancing their application in market design and beyond.

%https://www.nature.com/articles/s41599-024-03611-3 
%https://aws.amazon.com/blogs/hpc/llms-the-new-frontier-in-generative-agent-based-simulation/
%https://auction-arena.github.io/ focuses on the planning and iterate capability the bidding; the bidding behavior also has some rules. 
%discuss information rent 

%contribution: 
%- first framework for llm agent on market design with information disclosure 
%- the framework can be flexible to extend to iteration and adaptive
%- fully rely the llm reasoning to bid withset any rule like in 
%- for auction theory: gap the theoratical foundation with empirical results and also provide reasoning with llm knowledge. This also prove the usability of llm agents in running large scale simulation and help shed insights. 

\bibliography{aaai25}
\newpage
\appendix
\onecolumn
\section{Appendix}
\subsection{Details of the Framework}
\subsubsection{Auction and Signaling Process}
In this section, we introduce the detailed auction and signaling process used in our framework:
\begin{enumerate}
    \item \textbf{Auction Context Initialization}: In this initial phase, the auctioneer gathers all relevant information for the current auction round.
    \item \textbf{Information Disclosure}: The auctioneer discloses information to bidders according to the pre-defined strategy. This process includes public information, such as the number of participants or common priors, and private signals tailored to each bidder. 
    \item \textbf{Signal Reception and Belief Update}: Upon receiving the disclosed information, bidders analyze received information to update their beliefs about their true valuations and adapt bidding decision strategically. This process is operationalized through LLMs. 
    \item \textbf{Bid Generation and Submission}: Guided by their updated beliefs, bidders generate their bids for the current auction round. This process is operationalized also through LLMs. 
    \item \textbf{Bid Collection and Auction Execution}: The auctioneer collects all submitted bids and runs the auction based on the configured mechanism. 
    \item \textbf{Outcome Announcement} After executing the auction, the auctioneer publicly announces the results, potentially including the winner’s identity and the price paid. 
\end{enumerate}

\subsection{Prompt Inputs to Bidder Agents}

\begin{bidderbox}{Information Structure}
\begin{enumerate}
    \item \textbf{Default public information}: \texttt{You are a rational bidder in a second-price auction. The auction is one-time: you submit the bid and will receive the results. As a prior, you don't know your true value towards the current item. You only know your true-value falls between [0, 1]. Auctioneer may give you more information. The value of the item for each bidder is independently drawn from a uniform distribution over [0, 1] and the auction will consist of 10 bidders, including you. This is common knowledge to all bidders. However, your exact value is private and known only to you, based on signals or disclosed information provided by the auctioneer.}
    \item \textbf{Private signaling}: 
    \begin{itemize}
        \item Full Disclosure: \texttt{Your true value towards this current auctioned item is \{current average value\}}
        \item Tiered Disclosure without Average Tier Value: \texttt{Your true value towards this current auctioned item is not disclosed due to your value is being in the \{POOL LEVEL\} value tier. This indicates your value towards this item is \{POOL LEVEL\}er than some of other bidders, but the exact value will remain unknown.}
        \item Tiered Disclosure with Average Tier Value: \texttt{Your true value towards this current auctioned item is not disclosed due to your value is being in the \{POOL LEVEL\} value tier. This indicates your value towards this item is \{POOL LEVEL\}er than some of other bidders, but the exact value will remain unknown. The average value of all bidders in the same tier with you is \{TIER AVG VALUE\}}.
        \item Randomized Pool: \texttt{You have no information about your true value.}
    \end{itemize}
\end{enumerate}
\end{bidderbox}
\begin{bidderbox}{Bidding Instruction}
\texttt{Based on this information, submit a bid between 0 and 1, and provide the following details in your response: name: your name as a bidder. bid: Your bid for this round (a value between 0 and 1). estimated value: Your estimate of your true value based on the information received. explanation: A brief explanation of how you decided at your bid. Please return your response in the provided format}
\end{bidderbox}
\subsection{Simulation Variants}
Table \ref{table_xp} in illustrates the specific configurations in our simulation  design.
\begin{table*}[h]
\centering
%\resizebox{.95\columnwidth}{!}{
\begin{tabular}{l|l}
\textbf{Factor}                     & \textbf{Variations}                                                                                              \\ \hline
\textbf{Number of Bidders}          & 10                                                                                                                        \\ 
\textbf{Value Disclosure Strategy}   & Full Disclosure, Pool-High, Pool-Low, Randomized Pooling                                                                                        \\ 
\textbf{Threshold Quantiles}         & 0.2, 0.4, 0.6, 0.8                                                                                                         \\ 
\textbf{Pooled Information for Tiered Pooling} & Tier Only, Tier and Average Value Information                                                                                                                                                                                          \\ 
\textbf{Repetitions per Configuration} & 100                                                                                                                  
\end{tabular}
%}
\caption{Simulation Variations}
\label{table_xp}
\end{table*}

\subsection{Bidder's Reasoning}

\begin{bidderbox}{Examples of Bidder's Explanation on Bidding Decision}
\begin{enumerate}
    \item \textbf{Explanation under Full Disclosure}: \texttt{In a second-price auction, the optimal strategy is to bid your true value. This is because if you win, you will pay the second-highest bid, not your own. Therefore, bidding your true value maximizes your utility, ensuring you win the item if the second-highest bid is less than or equal to your true value. Given my true value is 0.3568638462861372, I will bid exactly this amount.}
    \item \textbf{Explanation under Tiered Pooling}: 
    \begin{itemize}
        \item Bid truthfully on the disclosed average value: \texttt{Given the information that my value is in the low value tier and the average value of bidders in this tier is 0.17774305518631073, I estimate my true value to be around this average. In a second-price auction, the optimal strategy is to bid your true value. Therefore, I will bid 0.17774305518631073, aligning my bid with my estimated true value..}
        \item Bid apart from the estimated value :\texttt{Given that my value is in the high value tier and the average value of this tier is 0.6610732108060862, I estimate my true value to be slightly above this average, at around 0.7. In a second-price auction, the optimal strategy is to bid your true value. However, since I don't know my exact value, I will bid the average of my tier, 0.6610732108060862, as it is a reasonable estimate of my true value.}
    \end{itemize}
    \item \textbf{Explanation under Randomized Pooling}: 
    \begin{itemize}
        \item Bid without any  information disclosed: \texttt{Without specific information about my true value, I assume the expected value of my true value is the midpoint of the distribution, which is 0.5. In a second-price auction, the optimal strategy is to bid your true value. Therefore, I bid 0.5, which is my estimated value based on the uniform distribution assumption.}
    \end{itemize}

\end{enumerate}
\end{bidderbox}

\end{document}